\documentclass[prd,preprint,tightenlines,floatfix,showpacs,preprintnumbers,nofootinbib]{revtex4}
\usepackage[dvips,final]{graphicx}
\begin{document}
    
\thispagestyle{empty}
\preprint{\hbox{CERN-TH-PH/2012-021}}

\title{Analytical eighth-order light-by-light QED contributions
from leptons with  heavier masses to the anomalous magnetic moment of
electron}

\author{A.~L.~Kataev}\
\email{kataev@ms2.inr.ac.ru}
\affiliation{Institute for Nuclear
Research of the Russian Academy of Sciences,  117312 Moscow, Russia}
\footnote{The preliminary version of the longer work with the same title 
was registered as CERN Preprint during the visit to Theory Division of CERN in January of 2012}

\vspace {10mm}
\begin{abstract}
The important consequences of the recent results of the numerical 
evaluations of eighth and tenth order QED contributions to 
the anomalous magnetic moment of electron are commented. The 
correctness of the results of the  numerical 
evaluation of new eighth order QED corrections to the electron anomaly 
are supported by the demonstration of their consistency 
 with the new analytical 
expressions for the QED contributions to 
$a_e$ from the diagrams with fourth-order light-by-light scattering 
muon and tau-lepton loops. The consistency  of the similar results
are demonstrated in the case of  eighth order massive dependent contribution 
to the muon anomalous magnetic moment,    
\end{abstract}

\maketitle

\label{sec:QED}
One of the most precise  at present  experimental results    in the modern
particle physics is the   measurement    of  the   electron
anomaly $a_e=(g_e-2)/2$ \cite{Hanneke:2008tm}, \cite{Hanneke:2010au},
which gives
\begin{equation}
a_e=1 159 652 180.73 (0.28) \times 10^{-12}~~~~~~[0.24ppb]~~~.
\label{experiment}
\end{equation}
The previous stage of   the  Standard Model  theoretical   prediction
\begin{equation}
\label{theory}
a_e^{th}=a_e^{QED} +a_e(hadrons)+a_e(weak)
\end{equation}
was summarized in the most
detailed review  on the subject \cite{Jegerlehner:2009ry}.
The perturbative  QED contribution to Eq.(\ref{theory})  is defined as
\begin{equation}
\label{expression}
a_e^{QED}=
A_1+A_2(m_e/m_{\mu})+A_2(m_e/m_{\tau})+A_3(m_e/m_{\mu}, m_e/m_{\tau})~~~.
\end{equation}
Up to recently  theoretical and phenomenological applications were
based  on
the following expression for its dominant   term $A_1$
\begin{eqnarray}
\label{A1}
A_1&=&\sum_{l=1}^{5}A_1^{(2l)}(\frac{\alpha}{\pi})^l=0.5\bigg(\frac{\alpha}{\pi}\bigg)
-0.32847896557919378 \dots \bigg(\frac{\alpha}{\pi}\bigg)^2\\
\nonumber
& +&
1.181241456587  \dots \bigg(\frac{\alpha}{\pi}\bigg)^3-
1.9144(35)\bigg(\frac{\alpha}{\pi}\bigg)^4+0.0(3.8)\bigg(\frac{\alpha}{\pi}\bigg)^5
\end{eqnarray}
where $l$ is the number of loops of the    Feynman diagrams,
which are contributing to  the corresponding perturbative QED
expression.
Three first coefficients, presented
in Eq.(\ref{A1}) in the numerical form,  were evaluated  analytically.
The first  term was calculated by Schwinger \cite{Schwinger:1948iu},
the second correction was  evaluated
by Petermann \cite{Petermann:1957hs} and Sommerfield
\cite{Sommerfield:1957zz}. This result was confirmed later  on by
Terentiev \cite{Terentiev}, who used   different technique.
The project of analytical evaluation of all 
three-loop QED contributions  to $a_e$   was completed
by Laporta and Remiddi \cite{Laporta:1996mq}. The cited
value for the four-loop  QED correction to $a_e$ was numerically obtained
in  Refs.\cite{Aoyama:2007dv}, \cite{Aoyama:2007mn} by Kinoshita
and   collaborators. The rough CODATA estimate   of the
coefficient of the 5-loop term in Eq.(\ref{A1}), namely
$A_1^{(10)}=\pm  2 |A_1^{(8)}|$
\cite{Mohr:2005zz}, 
gave  the idea what might be
theoretical uncertainties in the value of $a_e$ due to
unknown up to recently  total value of the    tenth-order
QED effects.
The project of their  direct numerical evaluation, started in 2005
by Kinoshita and Nio in Ref. \cite{Kinoshita:2005sm},
continued in the series of works of
Refs.~\cite{Aoyama:2008gy}- \cite{Aoyama:2012fc},
was successfully
completed in May of  2012 by Ayoama, Hayakawa, Kinoshita and Nio
\cite{Aoyama:2012wj}.

In general, the coefficient  $A_1^{(10)}$ is defined by the contribution
of 12672 diagrams, which were classified  into 32 gauge-invariant
subsets.  The result of evaluation of the
final, most complicated Set V,  together
with more detailed numerical calculation  of the eighth-order contribution
to $a_e$, was reported
recently  in Ref. \cite{Aoyama:2012wj}.
The long-expected expression for the tenth-order
massless contribution \cite{Aoyama:2012wj} is : 
\begin{equation}
A_1^{(10)}=9.16(58) \label{A10}
\end{equation}
while at the eight-order level the following
new results were obtained \cite{Aoyama:2012wj}:
\begin{eqnarray}
A_1^{(8)}&=&-1.9106(20) \label{A8} \\
a_e^{(8)}&=&-1.9097(20) \label{a8}~~.
\end{eqnarray}
The  $a_e^{(8)}$-term differs from Eq.(\ref{A8}) due to the
inclusion of complete  mass-dependent contributions,
 numerically evaluated in the process of the
works of  Refs.\cite{Aoyama:2008gy}-\cite{Aoyama:2012fc}. These
massive-dependent effects,   summarized in
in the numerical form
in Ref.\cite{Aoyama:2012wj}, read 
\begin{eqnarray}
A_2^{(8)}(m_e/m_{\mu})&=&9.222(66)\times 10^{-4} \label{emu} \\
A_2^{(8)}(m_e/m_{\tau})&=&8.24(12)\times 10^{-6} \label{etau}    \\
A_3^{(8)}(m_e/m_{\mu}, m_e/m_{\tau})&=&7.465(18)\times 10^{-7} \label{a38}
\end{eqnarray}
While obtaining these results   
the   CODATA-2010 report \cite{Mohr:2012tt} mass  ratios
$m_e/m_{\mu}=4.83633166(12)\times 10^{-3}$,
$m_{e}/m_{\tau}=2.87592(26)\times 10^{-4}$ ,
$m_{\mu}/m_{\tau}=5.94649(54)\times 10^{-2}$,
with the fixed value of $\tau$-lepton
pole mass $m_{\tau}=1776.82(16)$~{\rm MeV}
\footnote{Recently measured value of the  $\tau$-lepton 
mass
$m_{\tau}= 1776.69^{+0.17}_{-0.19}\pm 0.15~{\rm MeV}$  \cite{Eidelman:2011zzb}
should not change a lot  the uncertainties of these 
ratios.},  were used.  The results of Eq.(\ref{emu}), 
Eq.(\ref{etau})
and Eq.(\ref{a38}) are  presented in more detailed form in
Table I  of Ref.\cite{Aoyama:2012wj} for 12 gauge-invariant groups of
4-loop  diagrams,
contributing to  $A_2$ and $A_3$-terms of  Eq.(\ref{expression}).
The summary of massless and massive-dependent
contributions to $A_1^{(10)}$ and $A_2^{(10)}(m_e/m_{\mu})$
are  presented in Table II of Ref.\cite{Aoyama:2012wj}.
Note, that the sum of 10-th order massive dependent terms 
result in the small value of the overall massive correction $A_2^{(10)}(m_e/m_{\mu})=
-0.00382(39)$ \cite{Aoyama:2012wj}.

It should be stressed,  that the determination of the  concrete value of 
$A_1^{(10)}$- and $a_e^{(8)}$-terms  is of
real   importance. Indeed, prior the work of Ref.\cite{Aoyama:2012wj}
the scientific community  faced with the   unique case,  when
the  comparison of the low-energy experimental result of Eq.(\ref{experiment})
with the massless perturbative QED predictions of Eq.(\ref{A1}), supplemented 
with  analytically known from the works of 
Refs.\cite{Elend}- \cite{Passera:2006gc} massive-dependent forth and 
sixth-order contributions into $A_2$ and $A_3$-terms and with the 
well-known values for  $a_e(hadrons)$ and $a_e(weak)$ 
(see the review  \cite{Jegerlehner:2009ry}),  namely   
\begin{eqnarray}
\label{weak}
a_e(weak)&=&0.0297(5)\times 10^{-12} \\
a_e(hadrons)&=&1.671(19)\times 10^{-12}~~~,
\label{hadr}
\end{eqnarray}
led to  the  value of the inverse fine coupling  constant 
\begin{equation}
\alpha^{-1}=137.03599084(33)(39)
\label{alpha}
\end{equation}
with theoretical uncertainties $\pm 39\times 10^{-8}$,
comparable with experimental ones $\pm 33 \times 10^{-8}$.

The dominant contribution to the  theoretical error, namely
$\pm 30 \times 10^{-18}$, was  defined  by  the   CODATA
estimate of the coefficient of the $O(\alpha^5)$-term in Eq.(\ref{A1}).
The additional  sizable  uncertainty in Eq.(\ref{alpha})
 came  from the theoretical
error of the numerical evaluation   of the   eighth-order contributions
to Eq.(\ref{A1}) performed in    Refs.\cite{Aoyama:2007dv},
\cite{Aoyama:2007mn}. There were no additional theoretical  
errors in the   sum of cited above analytically evaluated   
forth and sixth-order 
massive-dependent  contributions 
$A_2$ and $A_3$-terms.  Their total 
contribution   can be extracted from the summary part of 
Ref.\cite{Passera:2006gc} and reads:
\begin{eqnarray}
A_2(m_e/m_{\mu})& =&
5.19738667(26)\times 10^{-7} \bigg(\frac{\alpha}{\pi}\bigg)^2 -
7.37394155(27)\times 10^{-6}\bigg(\frac{\alpha}{\pi}\bigg)^3
\label{A2mu} \\
A_2(m_e/m_{\tau})&=&
1.83798(34)\times 10^{-9}\bigg(\frac{\alpha}{\pi}\bigg)^2 -
6.5830(11)\times 10^{-8}\bigg(\frac{\alpha}{\pi}\bigg)^3
\label{A2tau} \\
A_3(m_e/m_{\mu}, m_e/m_{\tau})& =&
0.1909(1)\times 10^{-12} \bigg(\frac{\alpha}{\pi}\bigg)^3 ~~.
\label{A3}
\end{eqnarray}
The errors in Eq.(\ref{A2mu})-Eq.(\ref{A3}) are related 
to the indicated above uncertainties of the  CODATA-2010 values of the 
ratios of leptons masses. 

As to the uncertainties of the important 
contributions from  $a_e(hadrons)$ and $a_e(weak)$, they were 
not sensitive at the previous stage of comparing theoretical and experimental 
predictions for $a_e$.

The results of Ref. \cite{Aoyama:2012wj} and
Eq.(\ref{A10}) allowed to solve this  intriguing problem and to
make theoretical uncertainties in the analog of Eq.(\ref{alpha})
less important , than experimental ones.
Indeed, substituting    new eighth- and tenth-order 
QED effects of Ref.  \cite{Aoyama:2012wj} into  the the procedure of  
the comparison  with the precise experimental result of Eq.(\ref{experiment}),
the authors of  Ref.\cite{Aoyama:2012wj} obtained  
more  precise value of the inverse fine coupling constant:  
 \begin{equation}
\alpha^{-1}=137.035 999 1657 (68)(46)(24)(331)~~~[0.25ppb]~~~.
\end{equation}
Here the first and second errors are related to 
the uncertainties of the numerical evaluation of the 
eighth and tenth-order QED corrections. The third error 
is determined by the combined uncertainties of the 
hadronic and electroweak contributions to $a_e$, which start to manifest 
themselves at this more precise level of perturbative QED calculations, 
while the fourth and  hugest uncertainty is determined by the  experimental 
error in  Eq.(\ref{experiment}).

In view of the importance of the results of the complicated eight 
and tenth-order numerical calculations it is highly desirable to perform the independent 
cross-checks if not all of them, but at least of  some their  parts. 

In this work this problem is studied  by analyzing the consistency 
of the numerical results for the massive -dependent 
contributions into  $A_2^{(8)}(m_e/m_{\mu})$ and 
$A_2^{(8)}(m_e/m_{\tau})$-corrections to the electron anomaly, 
which are described by the subset of diagrams, formed by external  
light-by-light scattering muon and $\tau$-lepton subgraphs with  
extra virtual photon, propagating inside this 
subgraph. The corresponding 
analytical expressions for the leading term of heavy-mass expansions 
of these  contributions follow from   
the obtained in Ref.\cite{Boughezal:2011vw} 
result of analytical QCD calculations  of the hadronic light-by-light 
scattering contribution to the lepton  anomalous magnetic moments $a_e$ 
and $a_{\mu}$ with hadronic effects, modeled by the internal 
light-by-light-scattering quark loop, crossed by virtual gluon. Taking 
into account that in QED one has  $C_F=1$ and
$\alpha_{\overline{MS}}=\alpha(1+ O(\alpha^2))$ (see 
Ref.\cite{Broadhurst:1992za} for details) we
get the corresponding   
analytical contribution of the leading 
term in heavy lepton mass expansion  of the 
the eighth-order QED  correction to $a_e$ from this subset 
of  massive-dependent light-by-light scattering graphs.  
The results contain the contributions of the Riemann $\zeta$-functions
$\zeta_{\rm k}=\sum_{n=1}^{\infty}(1/n)^{\rm k}$
 and polylogarithmic functions
$\rm{a_k}=\rm{Li}_k(1/2)=\sum_{k=1}^{\infty}(1/2^k n^k)$ and read:
\begin{eqnarray}
\nonumber
&&A_2^{(8)}(X_{1+i}, lbl, NLO)\bigg(\frac{\alpha}{\pi}\bigg)^4 =\bigg( -\frac{473}{180}\zeta_2 \ln^{2}2+\frac{312}{405}\zeta_2\ln^{3}2-
\frac{42853}{2880}\zeta_4+\frac{5771}{360}\zeta_4\ln2 \\ \nonumber
&&+\frac{473}{1080}\ln^{4}2-\frac{52}{675}\ln^{5}2-
\frac{8477}{2700}+\frac{473}{45}a_4 +\frac{416}{45}a_5+\frac{34727}{2400}\zeta_3-
\frac{23567}{1440}\zeta_5 \bigg)\sum_{i=1}^{2}X_{1+i}^2\bigg(\frac{\alpha}{\pi}\bigg)^4~~~~.
\label{four}
\end{eqnarray}
where $X_{1+i}=(m_1/m_{1+i})$ with $i=1,2$
are the ratios  of the electron mass $m_1=m_e$ and the  muon mass  
$m_{2}=m_{\mu}$ or $\tau$-lepton  mass  $m_{3}=m_{\tau}$.
The abbreviations in the parenthesis labels  
the expressions for the
subsets of graphs, which contain  the  next-to-leading order ($NLO$) 
approximation of the light-by-light ($lbl$) scattering subgraph.

Substituting
the known  values for the transcendental and polylogarithmic functions
into Eq.(\ref{four}) we get:
\begin{equation}
\label{elnum}
A_2^{(8)}(X_{1+i},lbl,NLO)\bigg(\frac{\alpha}{\pi}\bigg)^4 = 1.77831 \dots \sum_{i=1}^{2} X_{1+i}^2
\bigg(\frac{\alpha}{\pi}\bigg)^4~~~.
\end{equation}

The CODATA-2010 values and errors of $X_1=m_e/m_{\mu}$ 
and $X_2=m_e/m_{\tau}$ are fixing  the numerical expressions 
for the terms we are interested in :   
\begin{eqnarray}
\label{a2mu}
A_2^{(8)}(X_2,lbl,NLO)\bigg(\frac{\alpha}{\pi}\bigg)^4&
= & 4.15948(21)\times 10^{-5}~ \bigg(\frac{\alpha}{\pi}\bigg)^4   \\
A_2^{(8)}(X_3,lbl,NLO)\bigg(\frac{\alpha}{\pi}\bigg)^4&=&1.47082(26)\times 10^{-7}~ \bigg(\frac{\alpha}{\pi}\bigg)^4
\label{a2tau}
\end{eqnarray}

The results of the  numerical calculations  for the 
similar  contributions are    presented 
in Table 1 of Ref. \cite{Aoyama:2012wj} and read 
\begin{eqnarray}
\label{A2MU}
A_2^{(8)}(X_2,lbl,NLO)\bigg(\frac{\alpha}{\pi}\bigg)^4&
= & 4.105(93)\times 10^{-5}~ \bigg(\frac{\alpha}{\pi}\bigg)^4  \\
A_2^{(8)}(X_3,lbl,NLO)\bigg(\frac{\alpha}{\pi}\bigg)^4&=&1.431(95)\times 10^{-7}~ \bigg(\frac{\alpha}{\pi}\bigg)^4
\label{A2TAU}~~~.
\end{eqnarray}
Taking into account 
the  related uncertainties one can observe 
good agreement with the numbers  of Eq.(\ref{a2mu}) and 
Eq.(\ref{a2tau}), which follow from analytical expression of Eq.(\ref{four}).
This can be considered as the strong check of the results 
of   computer  numerical calculations of eight-order contributions to $a_e$, 
summarized in Ref.\cite{Aoyama:2012wj}. Moreover, minor discrepancy 
between Eq.(\ref{a2tau}) and Eq.(\ref{A2TAU}) may indicate, that 
in the case of larger lepton mass ($m_{\tau}$) the exact numerical
results of Ref.\cite{Aoyama:2012wj} may be sensitive to taking into account 
higher terms of large mass expansion of the corresponding eight-order  
light-by-light-type diagrams, contributing to $a_e$. 

Let us have a look whether the similar feature is manifesting itself 
in the case of the comparison of  the results 
of analytical and numerical calculation  of the similar 
eight-order light-by-light scattering corrections to the anomalous  
magnetic moment of muon $a_{\mu}$.  
In this case the  
expression , analogous 
to Eq.(\ref{elnum}), reads  
\begin{equation}
A_2^{(8)}(X_4,lbl,NLO)\bigg(\frac{\alpha}{\pi}\bigg)^4 = 1.77831 \dots  X_{4}^2
\bigg(\frac{\alpha}{\pi}\bigg)^4~~~.
\end{equation}
Using the CODATA-10 value for $X_4=m_{\mu}/m_{\tau}$ we get 
\begin{equation}
A_2^{(8)}(X_4,lbl,NLO)\bigg(\frac{\alpha}{\pi}\bigg)^4 = 6.2888(11)\times 10^{-3}
\bigg(\frac{\alpha}{\pi}\bigg)^4~~~.
\label{taumu}
\end{equation}
This number is in good agreement with the result of the 
numerical eighth-order computer calculations, 
obtained in Ref. \cite{Aoyama:2012wk} in the process of completing 
numerical evaluation of tenth-order QED contributions to the muon anomaly 
$a_{\mu}$, namely with the result   
\begin{equation}
A_2^{(8)}(X_4,lbl,NLO)\bigg(\frac{\alpha}{\pi}\bigg)^4 = 
6.106(31)\times 10^{-3}\bigg(\frac{\alpha}{\pi}\bigg)^4~~~.
\label{TAUMU}
\end{equation}
However, as  in the case of  $\tau$-lepton 
light-by-light-scattering eighth order contributions
to $a_e$, the numerical comparison of the analytically-based result 
of Eq.(\ref{taumu}) with  the result of numerical calculations 
of Eq.(\ref{TAUMU}) seem to indicate the sensitivity  
to still unknown  higher terms of large mass expansions of 
the analytically evaluated massive-dependent Feynman graphs.

In any case our considerations demonstrates the reliability 
of the definite results of the complicated important  numerical QED
calculatiobns from  Ref.\cite{Aoyama:2012wj} and Ref.\cite{Aoyama:2012wk}.
More detailed considerations of  new 10-th QED  order results  for 
$a_{\mu}$ using the renormalization-group inspired studies  of 
Ref.\cite{Kataev:2006yh} may be done in future.

This work grew up from the  detailed studies of the results 
for  the eighth- and tenth-order QED contributions to the
anomalous magnetic moment of electron and muon, started   
in January of 2012 during the visit to   Theory Division of   CERN.
I am grateful to the members of Ph-Th Division of CERN for hospitality.
It is the pleasure to thank A.E. Dorokhov, M. Nio, M.Passera 
and T. Kinoshita  for discussions of various scientific subjects, 
related to this work.
This work is done within the scientific program of 
the  RFBR grants N 11-02-00112 and N 11-01-00182 and was supported
by  Grant NS-5590.2012.2.

\end{document}